\documentclass[12pt]{article}
\usepackage{cite}
\usepackage{amsmath,amssymb}
\begin{document}

\title{\bf{}Gauge invariant Lagrangian construction for massive bosonic mixed
symmetry higher spin fields}

\author{\sc I.L. Buchbinder\thanks{\tt joseph@tspu.edu.ru},
V.A. Krykhtin\thanks{\tt krykhtin@mph.phtd.tpu.edu.ru},
H. Takata\thanks{\tt takataxx@tspu.edu.ru}
\\[0.5cm]
\it Department of Theoretical Physics,\\
\it Tomsk State Pedagogical University,\\
\it Tomsk 634041, Russia
}

\date{}

\maketitle
\begin{abstract}
We develop the BRST approach to gauge invariant Lagrangian
construction for the massive mixed symmetry integer higher spin
fields described by the rank-two Young tableaux in arbitrary dimensional Minkowski space. The theory is
formulated in terms of auxiliary Fock space. No off-shell
constraints on the fields and the gauge parameters are imposed. The
approach under consideration automatically leads to a gauge
invariant Lagrangian for massive theory with all appropriate
St$\ddot{\rm u}$ckelberg fields. It is shown that all the
restrictions defining an irreducible representation of the Poincare
group arise from Lagrangian formulation as a consequence of the
equations of motion and gauge transformations. As an example of the
general procedure, we derive the gauge-invariant Lagrangian for
massive rank-2 antisymmetric tensor field containing the complete
set of auxiliary fields and gauge parameters.
\end{abstract}

\section{Introduction}

As known, in higher dimensions, $d>5$, the totally symmetric tensor
fields are not enough to cover all the irreducible representations
of the Poincare group. We should take into account the fields with
mixed symmetry of the indices as well. It is interesting that such
mixed symmetry fields naturally arise in context of superstring
theory. Therefore, a higher spin field theory in higher dimensions
should describe a mixed symmetry fields dynamics. Approaches to
Lagrangian formulation for the massless mixed symmetry fields in
Minkowski space were developed in refs. \cite{Ouvry}. Analogous
problems in AdS space were carefully analyzed in \cite{Brink}, where
the massless fields of generic symmetry type were studied and an
example of the Lagrangian for field corresponding to the three-cell
'hook' diagram was constructed (see also \cite{alk}).
Using dimensional reduction equations of motion in the 
St\"uckelberg
formalism for the massive 'hook' representation were obtained in \cite{0602225},
aspects of
Lagrangian formulation for massive higher spin mixed symmetry fields
were considered in refs. \cite{Zinoviev}.

In this note, we develop a general approach to gauge invariant
Lagrangian formulation for any massive bosonic mixed symmetry tensor
fields described by the rank-two Young tableaux in arbitrary dimensional Minkowsky space time. Our approach
is based on BRST-BFV method which earlier was applied for Lagrangian
construction for free totally symmetric massless and massive,
bosonic and fermionic higher spin fields in flat and in AdS spaces
\cite{BRST-HS,0505092}~\footnote{The massless higher spin
mixed symmetry fields were analyzed with help of BRST-BFV method in
refs. \cite{0101201}.}. Also the aspects of cubic interaction for
bosonic massless higher spin field were studied in this approach
\cite{BRST-int}.

The paper is organized as follows. In section 2 we briefly consider
a description of massive irreducible tensor representations of
$d$-dimensional Poincare group in terms of Fock space. To obtain the
irreducible tensors we should imposes the specific restrictions on
the fields. Then one introduces the standard creation and
annihilation operators and rewrites the above restrictions as the
operator constraints in Fock space. These operator constraints,
together with operators obtained from the constraints by Hermitian
conjugation, generate a closed algebra in terms of commutators. In
section 3 we construct, following refs. \cite{0505092} a new
representation of this algebra. In section 4 we derive the BRST-BFV
operator which then is used for constructing the Lagrangian and
gauge transformations. In section 5 we prove that equations defining
the irreducible representations are obtained from the found
Lagrangian after partial gauge fixing. In section 6 we consider two
examples. The first example shows that if we consider totally
symmetric tensor field corresponding to 1-row Young tableau, then
the method is reduced to the  case which is already studied in
\cite{0505092}. The second example deals with rank-2 antisymmetric
tensor field. We find the Lagrangian, the gauge transformations of
the fields, and the gauge transformation of the gauge parameters. In
section 7 we summarize the results and briefly discuss general mixed
symmetric theory.

\section{Irreducible mixed symmetric tensor representation of Poincare algebra}
In this paper we are going to construct a Lagrangian for the massive
tensor fields corresponding to a Young tableau with 2 rows ($s_1
\geqslant s_2$)
\begin{eqnarray}
\Phi_{\mu_1 \cdots \mu_{s_1},\,\nu_1 \cdots \nu_{s_2}} (x)
&\longleftrightarrow&
\begin{tabular}{|l|l|l|l|ll}
  \hline
  $\mu_1$ & $\mu_2$ & $\cdots$ & $\cdots$
  &\multicolumn{2}{l|}{$\cdots$ \vline $\;\;\mu_{s_1}$}\\
  \hline
  $\nu_1$ & $\nu_2$ & $\cdots$ & $\nu_{s_2}$ &  & \\
  \cline{1-4}
\end{tabular}
\label{basic}
.
\end{eqnarray}
In the correspondence with given Young tableau, the tensor field is
symmetric with respect to permutation of each type of the
indices\footnote{The indices inside round brackets are to be
symmetrized.}
\begin{math}
\Phi_{\mu_1 \cdots \mu_{s_1},\,\nu_1 \cdots \nu_{s_2}} (x)
=
\Phi_{(\mu_1 \cdots \mu_{s_1}),\,(\nu_1 \cdots \nu_{s_2})} (x)
\end{math}
and that after symmetrization of all indices corresponding to the
first row with one index corresponding to the second row the field
vanishes
\begin{eqnarray}
\Phi_{(\mu_1 \cdots \mu_{s_1},\,\nu_1) \cdots \nu_{s_2}} (x)
=0.
\label{const4}
\end{eqnarray}
We consider irreducibility condition under the mass stability
subgroup of the Poincare group which results in the traceless
condition\footnote{we use the metric $\eta^{\mu \nu} = {diag} ( +,
-, -, \cdots, - )$}
\begin{eqnarray}
\eta^{\mu_1\mu_2}
\Phi_{\mu_1 \mu_2\cdots\mu_{s_1},\,\nu_1\cdots\nu_{s_2}}
=
\eta^{\nu_1\nu_2}
\Phi_{\mu_1\cdots\mu_{s_1},\,\nu_1\nu_2\cdots\nu_{s_2}}
=
\eta^{\mu_1\nu_2}
\Phi_{\mu_1\cdots\mu_{s_1},\,\nu_1\cdots\nu_{s_2}}
=0
\label{traceless}
.
\end{eqnarray}
For irreducibility under the Poincare group we impose the
Klein-Gordon equation and the transversality conditions
\begin{eqnarray}
&&
( \partial^2 + m^2 )
\Phi_{\mu_1\cdots\mu_{s_1},\,\nu_1\cdots\nu_{s_2}} ( x ) = 0
,
\label{KG}
\\
&&
\partial^{\mu_1}
\Phi_{\mu_1\cdots\mu_{s_1},\,\nu_1\cdots\nu_{s_2}} ( x ) =
\partial^{\nu_1} \Phi_{\mu_1
\cdots\mu_{s_1},\,\nu_1\cdots\nu_{s_2}} ( x ) = 0
\label{transversal}
.
\end{eqnarray}

To avoid the explicit manipulations with a big number of indices it
is convenient to introduce Fock space generated by creation and
annihilation operators
\begin{eqnarray}
[a_i^{\mu}, a_j^{+\nu} ] = - \eta^{\mu \nu} \delta_{ij},
\qquad
i,j=1,2.
\label{[]}
\end{eqnarray}
The number of pairs of creation and annihilation operators one
should introduce is determined by the number of rows in the
Young tableau corresponding to the symmetry of the tensor field.
Thus, unlike the totally symmetric field case \cite{0505092} we introduce
two pairs of such operators.

An arbitrary state vector in this Fock space has the form
\begin{eqnarray}
|\Phi\rangle
&=&
\sum_{s_1=0}^{\infty} \sum_{s_2=0}^{\infty}
\Phi_{\mu_1\cdots\mu_{s_1},\,\nu_1\cdots\nu_{s_2}}(x)\;
a_1^{+\mu_1} \cdots a_1^{+\mu_{s_1}} a_2^{+\nu_1}\cdots
a_2^{+\nu_{s_2}}
|0\rangle
\label{Phi>}
.
\end{eqnarray}
One sees that the symmetry properties of the coefficient functions
\begin{math}
%\Phi_{\mu_1 \cdots \mu_{s_1},\,\nu_1 \cdots \nu_{s_2}}(x)=
\Phi_{(\mu_1 \cdots \mu_{s_1}),\,(\nu_1 \cdots \nu_{s_2})}(x)
\end{math}
are
stipulated by the symmetry properties of the product of the
creation operators.
To get another restrictions
(\ref{const4})--(\ref{transversal}) on the coefficient functions
we introduce the following operators
\begin{align}
&
l_0 = - p^\mu p_\mu + m^2,
&&
l_i = a_i^{\mu} p_{\mu},
&&
l_{{ij}} = \frac{1}{2} a_i^{\mu} a_{j \mu}
%\label{lij}
&&
g_{12} = - a_1^{+\mu} a_{2 \mu}
\label{g12}
\end{align}
where $p_\mu=-i\partial_\mu$. One can check that restrictions
(\ref{const4})--(\ref{transversal}) are equivalent to
\begin{align}
&
g_{12}|\Phi\rangle=0\,,
&&
l_{ij} | \Phi\rangle=0\,,
&&
l_0|\Phi\rangle=0\,,
&&
l_i|\Phi\rangle=0
\label{constraints}
\end{align}
respectively.

Our purpose is to construct a Lagrangian on the base of Hermitian
BRST-BFV operator. Therefore the set of operators which is used for
constructing the BRST-BFV  operator must be closed under Hermitian
conjugation and form a closed algebra. Thus, first, to form a set of
operators which is closed under Hermitian conjugation we add to
operators (\ref{g12}) their Hermitian conjugate
\begin{align}
&
l_i^{+} = a_i^{+\mu} p_{\mu}\,,
&&
l_{ij}^+ =\frac{1}{2} a_i^{+\mu} a_{j \mu}^+\,,
&&
g_{12}^+ = - a_2^{+\mu} a_{1 \mu} \equiv g_{21}
\label{l+}
\end{align}
and second, we add to the set of operators all the operators which
are necessary to form a closed algebra
\begin{align}
&
m^2\,,
&&
g_{11} =   - a_1^{+\mu} a_{1 \mu} + \frac{d}{2}
\,,
&&
g_{22}=  - a_2^{+\mu} a_{2 \mu} + \frac{d}{2}
\label{m2}
\,.
\end{align}
Now we have the set of operators (\ref{g12}), (\ref{l+}), (\ref{m2})
which is closed under Hermitian conjugation and form an algebra\footnote{It is worth noting that the same algebra of operators 
 can be obtained by first quantization of a spinning particle model
\cite{0702033}. We are grateful to A. Waldron for drawing our attention on this work.}
which given in Table~\ref{table}.
\begin{table}
\begin{center}
\begin{tabular}{|l||c|c|c|c||c|c|c|}
  \hline
  $\left[\; \downarrow, \rightarrow \right]$  & $m^2$ & $l_0$  &
            $l_k$  &  $l_k^+$  &   $l_{kl}$  &  $l_{kl}^+$  &  $g_{kl}$ \\
  \hline
  \hline
  $m^2 = m^2$  & 0 & 0  & 0  & 0  & 0  & 0  & 0  \\
  \hline
  $l_0 = - p^2 + m^2$  & 0&  0  &  0  &  0  &  0  &  0  &  0 \\
  \hline
  $l_i = p^\mu a_{i\mu}$  & 0 & 0  &  0  &  $\delta_{ik} ( l_0 - m^2 )$  &
  0  &  $-\delta_{i(l}l_{k)}^+$  &  $\delta_{i(k}l_{l)}$  \\
  \hline
  $l_i^+ = p^\mu a_{i\mu}^+$  & 0&  0 & $\cdots$  &  0  &
      $\delta_{i(l}l_{k)} $  &  0  &  $-\delta_{i(l}l_{k)}^+$  \\
  \hline
  \hline
  $l_{ij} = \frac{1}{2} a_{i\mu} a_j^\mu$  & 0 & 0  & 0 & $\cdots$  &  0  &
  ${\textstyle\frac{1}{2}}(\delta_{k(i}g_{j)l}+\delta_{l(i}g_{j)k})$
   &  $2\delta_{k(i}l_{j)l}$  \\
  \hline
  $l_{ij}^{+} = \frac{1}{2} a_i^{\mu+} a_{j\mu}^+$ & 0 & 0  &
  $\cdots$   & 0  & $\cdots$  &  0  &  $2\delta_{l(i} l_{j)k}^+$  \\
  \hline
  $g_{ij} = - a_{i\mu}^+ a_j^\mu + \frac{d}{2} \delta_{ij}$ & 0  &  0  & $\cdots$  & $\cdots$  & $\cdots$  & $\cdots$  &  $g_{i l} \delta_{j k} - g_{k j} \delta_{i l}$  \\
  \hline
\end{tabular}
\end{center}
\caption{Algebra of the initial operators}\label{table}
\end{table}

One can show that a straightforward use of the BRST-BFV construction
as if all the operators were first class constraints doesn't lead to
the proper equations (\ref{constraints}) for any value of spin (see
e.g. \cite{0505092} for the cases of higher spin fields
corresponding to one-row Young tableau). This happens because among
the above Hermitian operators there are operators which are not
constraints ($m^2$, $g_{11}$, $g_{22}$ in the case under
consideration) and they produce more equations (in addition to
(\ref{constraints})) for the physical field (\ref{Phi>}). Thus we
must somehow get rid of these supplementary equations. A method of
avoiding such supplementary equations consists of constructing new
enlarged expressions for the operators of the algebra given in
Table~\ref{table} so that the hermitian operators which are not
constraints will be zero.

\section{New representation of the algebra}

According to the method developed in our previous papers
\cite{0505092}, in order to avoid the supplementary equations for
the basic field one should construct new expressions for the
operators of the algebra given in Table~\ref{table}. These new
expressions $L_a$ are sums of the initial operators $l_a$ and
additional parts $l_a'$, that is $l_a\to{}L_a=l_a+l_a'$. The
additional expressions are constructed from new (additional)
creation and annihilation operators and parameters of the theory
(the mass $m$ in the theory under consideration) and so commute with
the initial operators. The requirements on the additional parts are:
1)~$L_a$ must form an algebra $[L_a,L_b]\sim{}L_c$; 2)~the
additional parts corresponding to the operators which produce the
supplementary equations  on the basic field must either linearly
contain arbitrary parameters whose values shall be defined later or
give zero when they are added to the corresponding initial
operators.

Since the initial operators commute with the additional parts
$[l_a,l_b']=0$ and since the algebra of the initial operators is a
Lie algebra then as was shown in \cite{0505092} the additional parts
must form the same algebra as the initial operators (see
Table~\ref{table}). Explicit expressions for the additional parts
satisfying a given algebra can be constructed using the method
adopted in \cite{Verma}, \cite{0505092}.

According to this method we introduce six pairs of bosonic
creation and annihilation operators with the standard commutation
relations
\begin{eqnarray}
[b_{11}, b_{11}^+]=[b_{22}, b_{22}^+]=[b_{12}, b_{12}^+]=[b, b^+]=
[b_{1}, b_{1}^+]=[b_{2}, b_{2}^+]
=1
\end{eqnarray}
and then derive explicit expressions for the additional
parts\footnote{ There are differences from paper\cite{0101201}
because of different definitions of generators (\ref{g12}).}
\begin{align}
\label{l0'}
&
l_0'=0\,,
&&
l_i'=mb_i\,,
&&
l_i^{\prime+}=mb_i^+\,,
&&
l_{ij}^{\prime+} = - \frac{1}{2} b_i^+ b_j^+ +b_{i j}^+
\,,
&&
m^{\prime2} = - m^2\,,
\end{align}
\begin{eqnarray}
l_{11}'
&=&
( h_1 - N_G ) b_{11} + \frac{1}{2} b^+ b_{12}
+ ( N_{11} + N_{12} ) b_{11} + \frac{1}{4} b_{22}^+ b_{12}^2
-\frac{1}{2}\,b_1^2
\,,
\\
l_{22}'
&=&
( h_2 + N_G ) b_{22} + \frac{1}{2} ( h_1 - h_2 - N_G )b b_{12}
+ ( N_{22} + N_{12} ) b_{22} + \frac{1}{4} b_{11}^+ b_{12}^2
-\frac{1}{2}\,b_2^2
\,,
\\
2 l_{12}'
&=&
( h_1 - h_2 - N_G ) b b_{11} + b^+ b_{22}
+\frac{1}{2} ( h_1 +h_2 ) b_{12}
\nonumber
\\
&&
\hspace*{5em}{}
+ \frac{1}{2}( 2N_{11}+ 2 N_{22} + N_{12} ) b_{12}
+2 b_{12}^+ b_{11} b_{22}
-b_1b_2
\,,
\\
g'_{11}
&=&
h_1 - N_G + 2 N_{11} + N_{12}+b_1^+b_1+\frac{1}{2}
\,,
\\
g_{22}'
&=&
h_2 + N_G + 2 N_{22} + N_{12}+b_2^+b_2+\frac{1}{2}
\,,
\\
g_{12}'
&=&
( h_1 - h_2 - N_G ) b + b_{11}^{\dagger} b_{12}
+2b_{12}^{\dagger} b_{22}+b_1^+b_2
\,,
\\
g_{21}'
&=&
b^+ + b_{22}^{\dagger} b_{12} +2 b_{12}^{\dagger}b_{11}
+b_2^+b_1
\label{g21'}
\,.
\end{eqnarray}
where
\begin{eqnarray}
N_{ij} = b_{ij}^+\; b_{ij},
&\qquad&
N_G = b^+ b.
\end{eqnarray}

The found additional parts possess all the necessary properties
described at the beginning of the present section. The enlarged
operators
\begin{align}
&
L_0=l_0+l_0',
&&
L_i=l_i+l_i',
&&
L_{ij}=l_{ij}+l_{ij}',
&&
G_{ij}=g_{ij}+g_{ij}',
\\
&
M^2=m^2+m^{\prime2},
&&
L_i^+=l_i^++l_i^{\prime+},
&&
L_{ij}^+=l_{ij}^++l_{ij}^{\prime+},
&&
G_{12}^+\equiv G_{21}
\end{align}
form an algebra (which coincides with the
algebra given in Table~\ref{table}) and the additional parts
$g_{11}'$, $g_{22}'$ corresponding to the operators $g_{11}$,
$g_{22}$ which are not constraints contain linearly arbitrary
parameters $h_1$, $h_2$ respectively or give zero\footnote{In what
follows we forget about operator $M^2$ since its enlarged
expression is zero.} $M^2=m^2+m^{\prime2}=0$.

It is easy to see that the additional parts do not possess the
needed Hermitian conjugation properties
\begin{align}
&
(l_{ij}')^+\neq l_{ij}^{\prime+}
\,,
&&
(g_{12}')^+\neq g_{21}'
\,.
\end{align}
The violation occurs in the $(b_{ij}^+,b^+)$ sector of the
auxiliary Fock space. Therefore to restore the Hermitian
conjugation properties we change the scalar product in this
sector of the Fock space
\begin{eqnarray}
\langle\Psi_1|\Psi_2\rangle_{new}&=&\langle\Psi_1|K|\Psi_2\rangle
\end{eqnarray}
with some operator $K$ to be defined.
This operator must restore the Hermitian conjugation
properties
\begin{eqnarray}
\langle\Psi_1|Kl_{ij}'|\Psi_2\rangle
=\langle\Psi_2|Kl_{ij}^{\prime+}|\Psi_1\rangle^*
\,,
&\qquad&
\langle\Psi_1|Kg_{12}'|\Psi_2\rangle
=\langle\Psi_2|Kg_{21}'|\Psi_1\rangle^*
\,.
\end{eqnarray}
The operator $K$ can be written in the form (see e.g. \cite{0505092}
for a more detailed explanation)
\begin{eqnarray}
\label{K}
&&K = Z^+ Z,
\\
&&
Z = \sum_{n_{{ij}}, n_G}
  ( l_{11}^{\prime+} )^{n_{11}} ( l_{22}^{\prime+} )^{n_{22}}
  (l_{12}^{\prime+} )^{n_{12}} ( g_{12}^{\prime+} )^{n_G} |0 \rangle_V
\frac{1}{n_{11}!n_{22} !n_{12} !n_G !}
\langle 0| ( b_{11})^{n_{11}} ( b_{22} )^{n_{22}}
( b_{12} )^{n_{12}} ( b )^{n_G}
\,,
\end{eqnarray}
where the auxiliary state $|0\rangle_V$ satisfies the relations
\begin{eqnarray}
&&
l_{11}'|0\rangle_V=
l_{22}'|0\rangle_V=
l_{12}'|0\rangle_V=
g_{12}'|0\rangle_V=0\,,
\qquad
{}_V\langle0|0\rangle_V=1
\\
&&
{}_V\langle0|l_{11}^{\prime+}=
{}_V\langle0|l_{22}^{\prime+}=
{}_V\langle0|l_{12}^{\prime+}=
{}_V\langle0|g_{12}^{\prime+}=0\,.
\end{eqnarray}

Thus in this section we construct the additional parts
(\ref{l0'})--(\ref{g21'}) which satisfied all the requirements.

\section{Constructing BRST-BFV operator and Lagrangian}
According to our method we should find the BRST-BFV operator. Since
the algebra under consideration is a Lie algebra this operator can
be constructed in the standard way\footnote{Use of the BRST approach
to higher spin field theory in AdS space leads to problem of
constructing a BRST-BFV operator for non-linear algebras, see e.g.
\cite{bl}.}. First we introduce the ghost fields with the
anticommutation relations
\begin{eqnarray}
\{\eta_0, P_0\}=
\{\eta_{12}^+,P_{12} \}=
\{\eta_{12},P_{12}^+ \}=
\{\eta_{12}^{G +},P^G_{12} \}=
\{\eta_{12}^G, P_{12}^{G +} \} = 1
\\
\{ \eta_i^+, P_k \}=
\{ \eta_i, P^+_k \}=
\{ \eta^+_{{ii}}, P_{{kk}} \}=
\{ \eta_{{ii}},P_{{kk}}^+ \} =
\{ \eta^G_{{ii}}, P_{{kk}}^G \}=
\delta_{{ik}}
\end{eqnarray}
with ghost numbers $+1$ for $\eta$'s and $-1$ for $P$'s. Then the
the BRST-BFV operator has the form
\begin{eqnarray}
\tilde{Q}=
Q
+\left[
 \eta^G_{11} ( \sigma_1 + h_1 )
 - (  \eta^+_{11} \eta_{11} +  \eta^+_{12}\eta_{12}
+  \eta^{G+}_{12} \eta^G_{12} )  { P^G_{11}} + ( 1 \leftrightarrow 2 )
\right],
\qquad
\tilde{Q}^2=0
,
\label{tildeQ}
\end{eqnarray}
where, for the later use, we have decomposed $\tilde{Q}$ in terms of
$\eta_{11}^G$ and $\eta_{22}^G$ and their conjugate momenta.
In (\ref{tildeQ}) three operators have been introduced
\begin{eqnarray}
Q&=&
\eta_0 L_0
+\sum_i ( \eta_i^+ L_i + \eta_i L_i^+ )
+\sum_{i\leq j} ( \eta_{ij}^+ L_{ij}+\eta_{ij} L_{ij}^+)
+\eta_{12}^{G+} G_{12} + \eta_{12}^G G^+_{12}
\nonumber
\\
&&{}
+ \Big [ - \eta_1^+ \eta_1 P_0 + \eta_1^+ (  \eta_{11} P_1^+ +\frac{1}{2} \eta_{12} P_2^+ ) - \eta_1 (  \eta^+_{11} P_1 +\frac{1}{2} \eta^+_{12} P_2 ) \nonumber
\\
&&{}
+  \eta^{G+}_{12} \left( \eta^+_1 P_2 - \eta_2 P_1^+ + 2\eta^+_{11} P_{12} - 2\eta_{22} P_{12}^+ + \eta_{12}^+ P_{22} - \eta_{12} P_{11}^+ \right) \nonumber
\\
&&{}
- \frac{1}{2} \left(  \eta^+_{22} \eta_{12} + \eta^+_{12} \eta_{11} \right) P^G_{12}
+ ( 1 \leftrightarrow 2 ) \Big] \label{Q}
\end{eqnarray}
and
\begin{eqnarray}
\sigma_i +  h_i & =&  G_{ii} + ( \eta^+_i P_i - \eta_i P_{i}^+ )  + 2 ( \eta^+_{ii}P_{ii} - \eta_{ii} P_{ii}^+ ) + ( \eta^+_{12} P_{12} - \eta_{12}P_{12}^+ )  \nonumber
\\
&&{}\mp ( \eta^{G+}_{12} P^G_{12} - \eta^G_{12} P^{G +}_{12} )
\,,
\\
&&
{}
[\tilde{Q},\sigma_i]=0\,,
\qquad
[Q,\sigma_i]=0\,,
\qquad
[\sigma_i,\sigma_j]=0\,,
\label{s2}
\end{eqnarray}
where in $\mp$ the upper sign corresponds to $i=1$ and the lower
one corresponds to $i=2$.

Further, we choose a representation of the Hilbert space given
by the relations
\begin{eqnarray}
(\eta_i, \eta_{ii}, \eta_{12}, \eta_{12}^G,
P_0, P_i, P_{ii}, P_{12}, P_{12}^G, P_{ii}^G)|0\rangle=0
\end{eqnarray}
and suppose that the field vectors as well as the gauge parameters do
not depend on $\eta_{ii}^G$
\begin{eqnarray}
|\chi \rangle &=& \sum_n ( a_{1 \mu_1}^+ \cdots a_{1 \mu_{n_{a 1}}}^+ ) ( a_{2\nu_1}^+ \cdots a_{2 \nu_{n_{a 2}}}^+ )  \nonumber
\\
&&{}\times
( b_1^+ )^{n_{b 1}} ( b_2^+ )^{n_{b2}} ( b_{11}^+ )^{n_{b 11}} ( b_{22}^+ )^{n_{b 22}} ( b_{12}^+ )^{n_{b 12}} (b^+ )^{n_b}\nonumber
\\
&&{}\times
( \eta_0^+ )^{n_{f 0}} ( \eta_1^+ )^{n_{f 1}} ( P_1^+ )^{n_{p 1}} ( \eta_2^+)^{n_{f 2}} ( P_2^+ )^{n_{p 2}}\nonumber
\\
&&{}\times
( \eta_{11}^+ )^{n_{f 11}} ( P_{11}^+ )^{n_{p11}} ( \eta_{22}^+ )^{n_{f 22}} ( P_{22}^+ )^{n_{p 22}} ( \eta_{12}^+)^{n_{f 12}}
( P_{12}^+ )^{n_{p 12}} ( \eta_{12}^{G +} )^{n_{{fg} 12}} (P^{G+}_{12} )^{n_{{pg} 12}}
\nonumber
\\
&&{}\times
\chi ( x )_{\mu_1 \cdots \mu_{n_{a 1}} \nu_1
\cdots \nu_{n_{a 2}}n_{p 1} n_{p 2} n_{p 11} n_{p 22} n_{p
12}n_{{fp}12}}^{n_{b_1} n_{b_2} n_{b 11} n_{b 22} n_{b 12} n_b
n_{f 0} n_{f 1}n_{f 2} n_{f 11} n_{f 22} n_{f 12} n_{{fg} 12}}
|0\rangle
\,.
\label{chi}
\end{eqnarray}
The sum in (\ref{chi}) is taken over $n_{ai}$, $n_{bi}$,
$n_{bij}$, $n_b$ running from $0$ to infinity, and over the rest
$n$'s from $0$ to $1$.
Let us denote by $|\chi^k\rangle$ the state (\ref{chi}) with the
ghost number $-k$, i.e. $gh(|\chi^k\rangle)=-k$.
Thus the physical state having the ghost number zero is
$|\chi^0\rangle$, the gauge parameters having the ghost number
$-1$ is $|\chi^1\rangle$ and
so on.

Since the vectors (\ref{chi}) do not depend on $\eta_{ii}^G$ the
equation for the physical state $\tilde{Q}|\chi^0\rangle=0$ yields
three equations
\begin{eqnarray}
&&Q|\chi^{0} \rangle=0,
\label{EOM}
\\
&&
(\sigma_i +h_i)|\chi^{0} \rangle=0,
\qquad
i=1,2
\,.
\label{EOM'}
\end{eqnarray}
Equations (\ref{EOM}), (\ref{EOM'}) are compatible due to
(\ref{s2}).
Equations (\ref{EOM'}) present equations for possible values of
$h_i$
\begin{eqnarray}
h_i&=&-\left( n_i + \frac{d - 7 \pm 2}{2} \right)\,,
\label{h}
\qquad
n_1=0,\pm1,\pm2,\ldots
\qquad
n_2=0,1,2,\ldots
\end{eqnarray}
Let us denote the eigenvectors of $\sigma_i$ corresponding to the
eigenvalues $ n_i + \frac{d - 7 \pm 2}{2} $ as
$|\chi\rangle_{n_1,n_2}$. Thus we may write
\begin{eqnarray}
\sigma_i | \chi \rangle_{n_1,n_2}
=
\left( n_i + \frac{d - 7 \pm 2}{2} \right) | \chi \rangle_{n_1,n_2}
\label{state}
\,.
\end{eqnarray}
Then one can show that in order to construct Lagrangian for the
field corresponding to a definite Young tableau (\ref{basic})
the numbers $n_i$ must be equal to the numbers of the boxes in the
$i$-th row of the corresponding Young tableau,
i.e.  $n_i=s_i$.
Thus the state $|\chi\rangle_{s_1s_2}$ contains the physical
field (\ref{basic}) and all its auxiliary fields.
Let us fix some values of $n_i=s_i$.
Then one should substitute $h_i$ corresponding to the chosen $n_i$
(\ref{h}) into (\ref{tildeQ}), (\ref{EOM}).
Thus the equation of motion (\ref{EOM}) corresponding to the
field with given spin $(s_1,s_2)$ has the form
\begin{eqnarray}
Q_{n_1n_2}|\chi^0\rangle_{n_1n_2}=0.
\label{Q12}
\end{eqnarray}

Since the BRST-BFV operator $\tilde{Q}$ is nilpotent (\ref{tildeQ})
at any values of $h_i$ we have a sequence of reducible gauge
transformation\footnote{ Since there are six momentum ghosts $P^+$'s
having negative ghost number, which are Grassman odd operators, the
lowest ghost number of $|\chi\rangle $ is $-6$ when $|\chi\rangle
\sim P^+_1P^+_2P^+_{11}P^+_{22}P^+_{12}P^{G+}_{12} | \mbox{no
ghosts} \rangle$ . }
\begin{eqnarray}
\label{dx0}
\delta|\chi^0 \rangle_{n_1,n_2} =Q_{n_1,n_2}|\chi^{1}\rangle_{n_1,n_2}
\,,
&\qquad&
\delta|\chi^{1}\rangle_{n_1,n_2} =Q_{n_1,n_2}|\chi^{2}\rangle_{n_1,n_2}
\,,
\\
\ldots
&\ldots&
\ldots
\\
\label{dx5}
\delta|\chi^5 \rangle_{n_1,n_2} =Q_{n_1,n_2}|\chi^{6}\rangle_{n_1,n_2}
\,,
&\qquad&
\delta|\chi^6 \rangle_{n_1,n_2} =0
\,.
\end{eqnarray}
One can show that $Q_{n_1,n_2}$ is nilpotent when acting $|\chi
\rangle_{n_1,n_2}$
\begin{eqnarray}
Q_{n_1,n_2}^2|\chi \rangle_{n_1,n_2}\equiv0.
\end{eqnarray}
Thus we have obtained equation of motion (\ref{Q12}) of
arbitrary spin gauge theory with mixed symmetry in any
space-time dimension and its tower of reducible gauge
transformations (\ref{dx0})--(\ref{dx5}).

We next find a corresponding Lagrangian .
Analogously to the bosonic one row case \cite{0505092} one can
show that Lagrangian for fixed spin $(n_1,n_2)$ is defined up to
an overall factor as follows
\begin{eqnarray}
{\cal L}_{n_1,n_2} = \int d \eta_0 \; {}_{n_1,n_2}\langle \chi |K_{n_1,n_2} Q_{n_1,n_2}| \chi \rangle_{n_1,n_2} \label{L}
\end{eqnarray}
where the standard scalar product for the creation and
annihilation operators is assumed and the operator $K_{n_1n_2}$
is the operator $K$ (\ref{K}) where the following substitution
is done $h\to-(n_i+(d-7\pm2)/2)$.

In the next section we show that the constraints (\ref{constraints})
can indeed be reproduced from (\ref{Q12}) up to gauge
transformations (\ref{dx0})--(\ref{dx5}).

\section{Reproduction of the initial constraints}\label{GFixing}

Let us show that the equations of motion
(\ref{const4})--(\ref{transversal}) [or equivalently
(\ref{constraints})] can be obtained from (\ref{Q12}) after partial
gauge-fixing and removing the auxiliary fields by using a part of
the equations of motion.

\subsection{Gauge-fixing}
Let us consider the field $|\chi^k \rangle$ at some fixed values of
the spin $(n_1,n_2)$. In this section we will omit the subscripts
associated with the eigenvalues of the $\sigma_i$ operators
(\ref{state}). Then we extract dependence of $Q$ (\ref{Q}) on zero
ghosts $\eta_0$ and $P_0$
\begin{eqnarray}
Q
&=&
\eta_0 L_0 - \left(\eta_1^+ \eta_1+\eta_2^+ \eta_2 \right)P_0 + \Delta Q
\\
\Delta Q
&=&
\sum_{i = 1, 2} ( \eta_i^+ L_i + \eta_i L_i^+ )
+ \sum_{(ij) = 11, 12, 22}
  ( \eta_{ij}^+ L_{ij} +\eta_{ij} L_{ij}^+ )
+ \eta_{21}^G G_{12} + \eta_{12}^G G_{21}
\nonumber
\\
&&{}
+ \eta_1^+ ( \eta_{11} P_1^+ + \frac{1}{2} \eta_{12} P_2^+ )
- \eta_1 (\eta^+_{11} P_1 + \frac{1}{2} \eta^+_{12} P_2 )
\nonumber
\\
&&{}
+ \eta_2^+ ( \eta_{22} P_2^+ + \frac{1}{2} \eta_{12} P_1^+ )
- \eta_2 (\eta^+_{22} P_2 + \frac{1}{2} \eta^+_{12} P_1 )
\nonumber
\\
&&{}
+ \eta^{G +}_{12} ( \eta^+_1 P_2 - \eta_2 P_1^+ + 2 \eta^+_{11} P_{12} - 2\eta_{22} P_{12}^+ + \eta_{12}^+ P_{22} - \eta_{12} P_{11}^+ )
\nonumber
\\
&&{}
+ \eta^G_{12} ( \eta^+_2 P_1 - \eta_1 P_2^+ + 2 \eta^+_{22} P_{12} - 2\eta_{11} P_{12}^+ + \eta_{12}^+ P_{11} - \eta_{12} P_{22}^+ )
\nonumber
\\
&&{}
- \frac{1}{2} ( \eta^+_{22} \eta_{12} + \eta^+_{12} \eta_{11} ) P^G_{12}
- \frac{1}{2} ( \eta^+_{11} \eta_{12} + \eta^+_{12} \eta_{22} ) P^{G +}_{12}
\label{DQ}
\end{eqnarray}
and do the same for the fields and gauge parameters
\begin{eqnarray}
|\chi^k\rangle=|S^k\rangle +\eta_0|A^k\rangle.
\end{eqnarray}
Then the gauge transformations and equations of motion
(\ref{Q12})--(\ref{dx5}) can be rewritten as follows
\begin{align}
&
\delta |S^{k - 1} \rangle = \Delta Q |S^k \rangle - (\eta_1^+ \eta_1+\eta_2^+ \eta_2) |A^k \rangle
&&
\delta |S^{-1} \rangle \equiv0,
\\
&
\delta |A^{k - 1} \rangle = L_0 |S^k \rangle - \Delta Q|A^k \rangle
&&
\delta |A^{-1} \rangle \equiv0.
\end{align}

Let us consider the lowest level gauge transformation
\begin{eqnarray}
\delta |S^{5} \rangle = \Delta Q|S^6 \rangle,
&\qquad&
\delta |A^{5} \rangle = L_0 |S^6 \rangle,
\end{eqnarray}
where due to the ghost number restriction one has used that
$|A^6\rangle\equiv0$. Extracting explicitly dependence of the gauge
parameters and of the operator $\Delta{}Q$ (\ref{DQ}) on
$\eta_{11}$, $P_{11}^+$ ghosts
\begin{eqnarray}
|\chi^k\rangle=|\chi^k_0\rangle+P_{11}^+|\chi^k_{1}\rangle,
&\qquad&
\Delta Q=\Delta Q_1+\eta_{11}T_1^++U_1P_{11}^+,
\end{eqnarray}
where $|\chi^k_0\rangle$, $|\chi^k_{1}\rangle$, $T_1^+$, $U_1$ do
not depend on $\eta_{11}$, $P_{11}^+$ we get the gauge
transformation of $|S^5_0\rangle$
\begin{eqnarray}
\delta |S^{5}_0 \rangle &=& T_1^+ |S^6_1 \rangle.
\label{dS5}
\end{eqnarray}
Here we have used that $|S^6_0\rangle\equiv0$ due to the ghost
number restriction. Since $T_1^+=b_{11}^++\ldots$ we can remove
dependence of $|S^5_0\rangle$ on $b_{11}^+$ using all the degrees of
freedom of $|S^6_1\rangle$. Thus, after the gauge fixing at the
lowest level of the gauge transformations we have conditions on
$|S_0^5\rangle$
\begin{eqnarray}
b_{11}|S^5_0\rangle=0
&\Longleftrightarrow&
b_{11}P_{11}^+|\chi^5\rangle=0.
\label{g5}
\end{eqnarray}

Let us turn to the next level of the gauge transformation.
Extracting explicit dependence of the gauge parameters and
$\Delta{}Q$ on $\eta_{11}$, $P_{11}^+$ and on $\eta_{22}$,
$P_{22}^+$ and using similar arguments as at the previous level of
the gauge transformation one can show that the gauge on
$|\chi^4\rangle$
\begin{eqnarray}
b_{11}P_{11}^+|\chi^4\rangle=0,
&\qquad&
b_{22}P_{22}^+P_{11}^+|\chi^4\rangle=0.
\end{eqnarray}
can be imposed. To obtain these gauge conditions all degrees of
freedom of the gauge parameters $|\chi^5\rangle$ restricted by
(\ref{g5}) must be used.

Applying a similar procedure one can obtain step by step
\begin{eqnarray}
b_{11}P_{11}^+|\chi^3\rangle=0, \qquad
b_{22}P_{22}^+P_{11}^+|\chi^3\rangle=0, \qquad
b_{12}P_{12}^+P_{22}^+P_{11}^+|\chi^3\rangle=0.
\end{eqnarray}
Then
\begin{eqnarray}
b_{11}P_{11}^+|\chi^2\rangle=0, \qquad
b_{22}P_{22}^+P_{11}^+|\chi^2\rangle=0, \qquad
b_{12}P_{12}^+P_{22}^+P_{11}^+|\chi^2\rangle=0, \qquad
b_{1}P_1^+P_{12}^+P_{22}^+P_{11}^+|\chi^2\rangle=0.
\end{eqnarray}
After this
\begin{align}
&
b_{11}P_{11}^+|\chi^1\rangle=0,
&&
b_{12}P_{12}^+P_{22}^+P_{11}^+|\chi^1\rangle=0,
&&
b_{2}P_2^+P_1^+P_{12}^+P_{22}^+P_{11}^+|\chi^1\rangle=0,
\\
& b_{22}P_{22}^+P_{11}^+|\chi^1\rangle=0, &&
b_{1}P_1^+P_{12}^+P_{22}^+P_{11}^+|\chi^1\rangle=0.
\end{align}
And finally we obtain gauge conditions on the field $|\chi^0\rangle$
\begin{align}
&
b_{11}P_{11}^+|\chi^0\rangle=0,
&&
b_{12}P_{12}^+P_{22}^+P_{11}^+|\chi^0\rangle=0,
&&
b_{2}P_2^+P_1^+P_{12}^+P_{22}^+P_{11}^+|\chi^0\rangle=0,
\label{G1}
\\
&
b_{22}P_{22}^+P_{11}^+|\chi^0\rangle=0,
&&
b_{1}P_1^+P_{12}^+P_{22}^+P_{11}^+|\chi^0\rangle=0,
&&
bP_{12}^{G+}P_2^+P_1^+P_{12}^+P_{22}^+P_{11}^+|\chi^0\rangle=0.
\label{G2}
\end{align}

Let us now turn to removing the auxiliary fields using the equations
of motion.

\subsection{Removing auxiliary fields by means of equations of motion}

First we decompose the fields $|S^0\rangle$ as follows
\begin{align}
&|S^0 \rangle = |S^0_0 \rangle + P_{11}^+ |S^0_1 \rangle,
&&|S^0_{000} \rangle = |S^0_{0000} \rangle + P_{1}^+ |S^0_{0001}\rangle,
\\
&|S^0_0\rangle = |S^0_{00} \rangle + P_{22}^+ |S^0_{01}
\rangle,
&&|S^0_{0000} \rangle = |S^0_{00000} \rangle + P_{2}^+
|S^0_{00001}\rangle,
\\
&|S^0_{00} \rangle = |S^0_{000} \rangle + P_{12}^+ |S^0_{001}
\rangle,
&&|S^0_{00000} \rangle = |S^0_{000000} \rangle + P_{12}^{G+} |S^0_{000001} \rangle
\end{align}
and do the same for $|A^0\rangle$
\begin{eqnarray}
|A^0 \rangle
&=&
P_{12}^{G+} |A^0_{000001} \rangle
+ P_{2}^+ |A^0_{00001} \rangle
+ P_{1}^+ |A^0_{0001} \rangle
+ P_{12}^+ |A^0_{001} \rangle
+ P_{22}^+ |A^0_{01} \rangle
+ P_{11}^+ |A^0_1 \rangle,
\end{eqnarray}
where the term independent of the ghost momenta is absent due to
the ghost number restriction.
We note that due to $gh(|S^0\rangle)=0$, $|S^0_{000000}\rangle$
can't depend on the ghost coordinates and as a consequence of
the gauge conditions (\ref{G1}), (\ref{G2})
$|S^0_{000000}\rangle=|\Phi\rangle$, with $|\Phi\rangle$ being
the physical field (\ref{Phi>}).

Then analogously to the fields we extract in $\Delta{}Q$
(\ref{DQ}) first dependence on $\eta_{11}$, $P_{11}^+$, next
dependence on $\eta_{22}$, $P_{22}^+$, and further  on
$\eta_{12}$, $P_{12}^+$, on $\eta_{1}$, $P_{1}^+$, on
$\eta_{2}$, $P_{2}^+$, and on $\eta_{12}^G$, $P_{12}^{G+}$
respectively.

Substituting these decompositions into the equation of motion
\begin{eqnarray}
l_0 |S^0 \rangle - \Delta Q|A^0 \rangle=0
\end{eqnarray}
and using the gauge conditions (\ref{G1}), (\ref{G2}) one can
show that first $|A^0_{000001}\rangle=0$, then
$|A^0_{00001}\rangle=0$, and so on till $|A^0_{1}\rangle=0$
which means that
\begin{eqnarray}
l_0|S^0\rangle=0,
&\qquad&
|A^0\rangle=0.
\label{E1}
\end{eqnarray}

Analogously we consider the second equation of motion
\begin{eqnarray}
\Delta Q |S^0 \rangle=0,
\end{eqnarray}
where $|A^0\rangle=0$ has been used.
After the same decomposition we conclude one after another that
\begin{eqnarray}
|S^0_{000001}\rangle=|S^0_{00001}\rangle=|S^0_{0001}\rangle=
|S^0_{001}\rangle=|S^k_{01}\rangle=|S^k_1\rangle=0.
\label{E2}
\end{eqnarray}
Eqs. (\ref{E1}) and (\ref{E2}) mean that all the auxiliary fields
vanish and as a result we have $|\chi^0\rangle=|\Phi\rangle$ and the
equations of motion (\ref{constraints}) are fulfilled.

Let us now turn to examples.

\section{Examples}
%%%%%%%%%%%%%%%%%%%%%%%%%%%%%%%
\subsection{Spin-$(s,0)$ totally symmetric field}
Let us consider the totally symmetric field corresponding to
spin-$(s,0)$. In this case we expect that our result will be reduced
to that considered in \cite{0505092}, where the totally symmetric
massive bosonic fields were considered. According to our procedure
we have $n_1=s$, $n_2=0$. One can show that if $n_2=0$ then in
(\ref{chi}) all the components related with the second row in the
Young tableau must be equal to zero, i.e.
\begin{eqnarray}
n_{a2}=n_{b2}=n_{b22}=n_{b21}=n_b=n_{f2}=n_{p2}
=n_{f22}=n_{p22}
=n_{f12}=n_{p12}
=n_{fg12}=n_{pg12}=0
\,.
\end{eqnarray}
Thus the state vector is reduced to
\begin{eqnarray}
|\chi \rangle
&=&
\sum_n ( a_{1 \mu_1}^+ \cdots a_{1 \mu_{n_{a 1}}}^+ ) ( b_1^+
)^{n_{b 1}} ( b_{11}^+ )^{n_{b 11}} ( \eta_0^+ )^{n_{f 0}} ( \eta_1^+ )^{n_{f
1}} ( P_1^+ )^{n_{p 1}} ( \eta_{11}^+ )^{n_{f 11}} ( P_{11}^+ )^{n_{p 11}}
\nonumber
\\
&&{}\times
\chi ( x )_{\mu_1 \cdots \mu_{n_{a 1}}0n_{p 1}0n_{p11}
000}^{n_{b_1}0n_{b11}000n_{f0}n_{f1}0n_{f11}000}
|0\rangle\,,
\label{x1}
\end{eqnarray}
which corresponds to that in \cite{0505092}. Then one can easily
show that equations (\ref{Q12}), (\ref{dx0}), (\ref{L}) with
$|\chi\rangle$ as in (\ref{x1}) reproduce the same relations as
those in \cite{0505092}.

%%%%%%%%%%%%%%%%%%%%%%%%%%%%%%%
\subsection{Rank-2 antisymmetric tensor field}
The next example is the simplest mixed symmetric case, that is,
spin-$(1,1)$ rank-2 totally antisymmetric tensor field.
In this example we denote all the gauge parameters by letters
with primes and all the gauge parameters of the second level
of the gauge transformations by letters with two primes.

\subsubsection{Lagrangian}
Let us decompose the state vector (\ref{chi}) having ghost number
zero\footnote{The states (\ref{chi}) in spin-$(1,1)$ case have the
lowest ghost number $-2$ due to restriction (\ref{state}) at
$n_1=n_2=1$.} and obeying (\ref{state}) at $n_1=n_2=1$ first in the
ghost fields and then in auxiliary creation operators $b_{11}^+$,
$b_{22}^+$, $b_{12}^+$, $b^+$
\begin{eqnarray}
|\chi^0\rangle_{1,1} &=&
|B\rangle_{1,1} + b^+|T_1\rangle_{2,0} + b_{12}^+|\phi_1\rangle_{0,0} + b_{11}^+b^+ |\phi_2\rangle_{0,0}
\nonumber
\\
&&{}
+P_1^+\eta_1^+b^+|\phi_3\rangle_{0,0}
+P_1^+\eta_2^+|\phi_4\rangle_{0,0}
+P_2^+\eta_1^+|\phi_5\rangle_{0,0}
\nonumber
\\
&&{}
+\eta_{12}^{G+}P_1^+|A_1\rangle_{1,0}
+\eta_{12}^{G+}P_{11}^+|\phi_6\rangle_{0,0}
+P_{12}^{G+}\eta_1^+|A_2\rangle_{1,0}
+P_{12}^{G+}\eta_{11}^+|\phi_7\rangle_{0,0}
\nonumber
\\
\nonumber
&+&\eta_0\Big(
P_1^+ \left( |H\rangle_{0,1}+b^+|A_3\rangle_{1,0} \right)
+P_2^+|A_4\rangle_{1,0}
+P_{11}^+b^+|\phi_8\rangle_{0,0}
+P_{12}^+|\phi_{10}\rangle_{0,0}
\nonumber
\\
&&{}
+P_{12}^{G+} \left(|T_2\rangle_{2,0} + b_{11}^+|\phi_9\rangle_{0,0} \right)
+P_{12}^{G+}P_1^+\eta_1^+|\phi_{11}\rangle_{0,0}
\Big)\,,
\label{x0}
\end{eqnarray}
where states $|\cdot\rangle$ still depend on
$a_{1,\mu}^+$, $a_{2,\mu}^+$, $b_1^+$, $b_2^+$
 and are expanded as follows
\begin{eqnarray}
|B\rangle_{1,1} &=&
- a_1^{+\mu}a_2^{+\nu} |0\rangle B_{\mu\nu}(x)
-i  b_1^+  a_2^{+\mu} |0\rangle H_{(2) \mu} (x)
-i  b_2^+ a_1^{+\mu} |0\rangle A_{(7) \mu} (x)
+  b_1^+ b_2^+ |0\rangle \phi_{(19)}(x)
\\
|T_{i}\rangle_{2,0} &=&
- a_1^{+\mu}a_1^{+\nu} |0\rangle T_{(i)(\mu\nu)}(x)
-i  b_1^+  a_1^{+\mu} |0\rangle A_{(4+i) \mu} (x)
+( b_1^+)^2 |0\rangle \phi_{(16+i)}(x)
\\
|H\rangle_{0,1} & = &
-i a_2^{+\mu} |0\rangle H_\mu (x)
+ b_2^+ |0\rangle \phi_{(16)} (x)
\\
|A_{i}\rangle_{1,0} & = &
-i a_1^{+\mu} |0\rangle A_{(i)\mu} (x)
+ b_1^+ |0\rangle \phi_{(11+i)} (x)
\\
|\phi_{i}\rangle_{0,0} &=&
|0\rangle \phi_{(i)}(x)
\end{eqnarray}

Then the relation (\ref{L}) gives the following Lagrangian
\begin{eqnarray}
{\cal L}_{(1, 1)}
&=&
B_{\mu \nu} (\partial^2+m^2) B^{\mu \nu}
- 4 T_{(2) \left(\mu \nu \right)} B^{\mu \nu}
- 4 T_{(1) \left(\mu \nu \right)} (\partial^2+m^2) T_{(1)}^{\mu \nu}
+ 8 T_{(1) \left(\mu \nu \right)} T_{(2)}^{\mu \nu}
\nonumber
\\
&&{}
+ 2 A_{(4) \mu} \partial_{\nu} B^{\mu \nu}
+ 2 H_{\nu} \partial_{\mu} B^{\mu \nu}
- 4 A_{(1) \mu} \partial_{\nu} T_{(2)}^{\mu \nu}
- 8 A_{(3) \mu} \partial_{\nu} T_{(1)}^{\mu \nu}
+\phi_{(10)} B_{\mu}^{\mu}
- 2\phi_{(6)} T_{\mu (2)}^{\mu}
\nonumber
\\
&&{}
- 4 \phi_{(8)} T_{\mu (1)}^{\mu}
- 2 A_{(1) \mu} (\partial^2+m^2) A_{(2)}^{\mu}
+ 2 A_{(5) \mu} (\partial^2+m^2) A_{(5)}^{\mu}
- A_{(7) \mu} (\partial^2+m^2) A_{(7)}^{\mu}
\nonumber
\\
&&{}
- H_{(2) \mu} (\partial^2+m^2) H_{(2)}^{\mu}
- 2m A_{(1) \mu} A_{(6)}^{\mu}
- 4 A_{(2) \mu} A_{(3)}^{\mu}
+ 2 A_{(2) \mu} A_{(4)}^{\mu}
+ 2 A_{(2) \mu} H^{\mu}
\nonumber
\\
&&{}
+ 2 A_{(3) \mu} A_{(3)}^{\mu}
- 4mA_{(3) \mu} A_{(5)}^{\mu}
- A_{(4) \mu} A_{(4)}^{\mu}
+ 2mA_{(4) \mu} A_{(7)}^{\mu}
- 4 A_{(5) \mu} A_{(6)}^{\mu}
+ 2 A_{(6) \mu} A_{(7)}^{\mu}
\nonumber
\\
&&{}
+ 2 A_{(6) \mu} H_{(2)}^{\mu}
- H_{\mu} H^{\mu}
+ 2mH_{\mu} H_{(2)}^{\mu}
+ 2 \phi_{(11)} \partial_{\mu} A_{(1)}^{\mu}
+ 4 \phi_{(3)} \partial_{\mu} A_{(3)}^{\mu}
- 2 \phi_{(4)} \partial_{\mu} A_{(4)}^{\mu}
\nonumber
\\
&&{}
+ 4 \phi_{(14)} \partial_{\mu} A_{(5)}^{\mu}
+ 2 \phi_{(12)} \partial_{\mu} A_{(6)}^{\mu}
- 2 \phi_{(16)} \partial_{\mu} A_{(7)}^{\mu}
- 2 \phi_{(5)} \partial_{\mu} H^{\mu}
- 2 \phi_{(15)} \partial_{\mu} H_{(2)}^{\mu}
\nonumber
\\
&&{}
+ \frac{5 - d}{4} \phi_{(1)} (\partial^2+m^2) \phi_{(1)}
- 2 \phi_{(1)} (\partial^2+m^2) \phi_{(2)}
+ (d - 1) \phi_{(2)} (\partial^2+m^2) \phi_{(2)}
\nonumber
\\
&&{}
- 2 \phi_{(1)} (\partial^2+m^2) \phi_{(2)}
+ (d - 1) \phi_{(2)} (\partial^2+m^2) \phi_{(2)}
+ 2 \phi_{(3)} (\partial^2+m^2) \phi_{(3)}
\nonumber
\\
&&{}
- 2 \phi_{(4)} (\partial^2+m^2) \phi_{(5)}
+ 2 \phi_{(6)} (\partial^2+m^2) \phi_{(7)}
+ 2 \phi_{(12)} (\partial^2+m^2) \phi_{(13)}
\nonumber
\\
&&{}
- 4 \phi_{(17)} (\partial^2+m^2) \phi_{(17)}
+ \phi_{(19)} (\partial^2+m^2) \phi_{(19)}
+ 2 \phi_{(1)} \phi_{(8)}
+ (d - 3) \phi_{(1)} \phi_{(9)}
\nonumber
\\
&&{}
+ \frac{d - 5}{2} \phi_{(1)} \phi_{(10)}
+ 2 (1 - d) \phi_{(2)} \phi_{(8)}
+ 2 (3 - d) \phi_{(2)} \phi_{(9)}
+ 2 \phi_{(2)} \phi_{(10)}
- 4 \phi_{(3)} \phi_{(8)}
\nonumber
\\
&&{}
- 4 \phi_{(3)} \phi_{(11)}
+ 4m \phi_{(3)} \phi_{(14)}
+ \phi_{(4)} \phi_{(10)}
+ 2 \phi_{(4)} \phi_{(11)}
- 2m \phi_{(4)} \phi_{(15)}
+ \phi_{(5)} \phi_{(10)}
\nonumber
\\
&&{}
+ 2 \phi_{(5)} \phi_{(11)}
- 2m \phi_{(5)} \phi_{(16)}
+ (3 - d) \phi_{(6)} \phi_{(9)}
+ \phi_{(6)} \phi_{(10)}
- 2 \phi_{(6)} \phi_{(11)}
- 2 \phi_{(6)} \phi_{(18)}
\nonumber
\\
&&{}
+ 4 \phi_{(7)} \phi_{(8)}
- 2 \phi_{(7)} \phi_{(10)}
- 4 \phi_{(8)} \phi_{(17)}
+ \phi_{(10)} \phi_{(19)}
+ 2m \phi_{(11)} \phi_{(12)}
+ 4m \phi_{(12)} \phi_{(18)}
\nonumber
\\
&&{}
+ 4 \phi_{(13)} \phi_{(14)}
- 2 \phi_{(13)} \phi_{(15)}
- 2 \phi_{(13)} \phi_{(16)}
- 2 \phi_{(14)} \phi_{(14)}
+ 8m \phi_{(14)} \phi_{(17)}
+ \phi_{(15)} \phi_{(15)}
\nonumber
\\
&&{}
- 2m \phi_{(15)} \phi_{(19)}
+ \phi_{(16)} \phi_{(16)}
- 2m \phi_{(16)} \phi_{(19)}
+ 8 \phi_{(17)} \phi_{(18)}
- 4 \phi_{(18)} \phi_{(19)}
\,.
\label{L(1,1)}
\end{eqnarray}

Let us turn to the gauge transformations.

\subsubsection{The gauge transformations for the fields}

Now we decompose the vector (\ref{chi}) obeying the (\ref{state}) at
$n_1=n_2=1$ and having ghost number $-1$ as follows
\begin{eqnarray}
|\chi^{1}\rangle_{1,1} &=&
P_1^+ \left( |H^\prime \rangle_{0,1}+b^+|A_1^\prime \rangle_{1,0} \right)
+P_2^+|A_2^\prime \rangle_{1,0}
+P_{11}^+b^+|\phi_1^\prime \rangle_{0,0}
+P_{12}^+|\phi_{2}^\prime \rangle_{0,0}
\nonumber
\\
&&{}
+P_{12}^{G+} \left(|T_1^\prime \rangle_{2,0} + b_{11}^+|\phi_3^\prime \rangle_{0,0} \right)
+P_{12}^{G+}P_1^+\eta_1^+|\phi_{4}^\prime \rangle_{0,0}
\nonumber
\\
&+&\eta_0 \Big(
P_1^+P_2^+|\phi_5^\prime \rangle_{0,0}
+P_1^+P_{12}^{G+}|A_3^\prime \rangle_{1,0}
+P_{11}^+P_{12}^{G+}|\phi_6^\prime \rangle_{0,0}
\Big)\,,
\label{x-1}
\end{eqnarray}
where
\begin{eqnarray}
|T^\prime\rangle_{2,0} &=&
- a_1^{+\mu}a_1^{+\nu} |0\rangle T^\prime_{\mu\nu}(x)
-i  b_1^+  a_1^{+\mu} |0\rangle A^\prime_{(4) \mu} (x)
+( b_1^+)^2 |0\rangle \phi^\prime_{(11)}(x)
\\
|H^\prime\rangle_{0,1} & = &
-i a_2^{+\mu} |0\rangle H^\prime_\mu (x)
+ b_2^+ |0\rangle \phi^\prime_{(10)} (x)
\\
|A^\prime_{i}\rangle_{1,0} & = &
-i a_1^{+\mu} |0\rangle A^\prime_{(i)\mu} (x)
+ b_1^+ |0\rangle \phi^\prime_{(6+i)}(x)
\\
|\phi^\prime_{i}\rangle_{0,0} &=&
|0\rangle \phi^\prime_{(i)}(x)
\,.
\end{eqnarray}
Substituting (\ref{x0}), (\ref{x-1}) in the left relations
(\ref{dx0}) we obtain gauge transformations for the fields
\begin{eqnarray}
\delta B_{\mu\nu}
&=&
2 {T^\prime_{\mu  \nu }}   + {\partial_{\mu }}  {H^\prime_{\nu }} + {\partial_{\nu }}  {A^\prime_{{(2)} \mu }}
-  \frac{{{\eta }_{\mu  \nu }}}{2} {\phi^\prime }_{(2)}
\\
\delta T_{{(1)}\mu\nu} &=&{T^\prime_{\mu  \nu }} + {\partial_{(\mu }}  {A^\prime_{(1)\nu) }} - \frac{{{\eta }_{\mu  \nu }}}{2}{{\phi^\prime }_{(1)}}
\\
\delta T_{{(2)}\mu\nu}&=& {(\partial^2+m^2)}  {T^\prime_{\mu  \nu }}    -{\partial_{(\mu }}  {A^\prime_{{(3)} \nu) }} + \frac{{{\eta }_{\mu  \nu }}}{2} {\phi^\prime }_{(6)}
\\
\delta A_{{(1)}\mu} &=& -2  {A^\prime_{(1)\mu }} +      {A^\prime_{{(2)} \mu }} + {H^\prime_{\mu }}
\\
\delta A_{{(2)}\mu} &=& -  2 \partial^{\nu }  {T^\prime_{\mu  \nu }} - m  {A^\prime_{(4)  \mu }} - {\partial_{\mu }}  {{\phi^\prime }_{(4)}} +  {A^\prime_{{(3)} \mu }}
\\
\delta A_{{(3)}\mu} &=&   {(\partial^2+m^2)}  {A^\prime_{(1)\mu }} + {A^\prime_{{(3)} \mu }}
\\
\delta A_{{(4)}\mu} &=&  -{\partial_{\mu }}  {{\phi^\prime }_{(5)}} + {(\partial^2+m^2)}  {A^\prime_{{(2)} \mu }} + {A^\prime_{{(3)} \mu }}
\\
\delta A_{(5)\mu} &=&  m  {A^\prime_{(1)\mu }} + {\partial_{\mu }}  {{\phi^\prime }_{(7)}} +      {A^\prime_{(4)  \mu }}
\\
\delta A_{(6)\mu}   &=&    -m  {A^\prime_{{(3)} \mu }} - {\partial_{\mu }}  {{\phi^\prime }_{(9)}} + {(\partial^2+m^2)}  {A^\prime_{(4)  \mu }}
\\
\delta A_{(7)\mu}  &=&   m  {A^\prime_{{(2)} \mu }} + {\partial_{\mu }}  {{\phi^\prime }_{(10)}} +  {A^\prime_{(4)  \mu }}
\\
\delta H_{\mu}  &=&   {\partial_{\mu }}  {{\phi^\prime }_{(5)}} + {(\partial^2+m^2)}  {H^\prime_{\mu }} + {A^\prime_{{(3)} \mu }}
\\
\delta H_{(2)\mu}  &=&  m  {H^\prime_{\mu }} + {\partial_{\mu }}  {{\phi^\prime }_{(8)}} +    {A^\prime_{(4)  \mu }}
\end{eqnarray}
\vspace*{-6ex}
\begin{eqnarray}
%\\
\delta \phi_{{(1)}} &=&  {{\phi^\prime }_{(2)}} + 2 {{\phi^\prime }_{(3)}}
\\
\delta \phi_{{(2)}}  &=& {{\phi^\prime }_{(1)}} + {{\phi^\prime }_{(3)}}
\\
\delta \phi_{{(3)}}&=& -  \partial^{\mu }  {A^\prime_{(1)\mu }} - m  {{\phi^\prime }_{(7)}} + {{\phi^\prime }_{(1)}} +   {{\phi^\prime }_{(4)}}
\\
\delta \phi_{{(4)}}   &=& -  \partial^{\mu }  {H^\prime_{\mu }} - m  {{\phi^\prime }_{(10)}} + \frac{{1}}{2} {\phi^\prime }_{(2)}+ {{\phi^\prime }_{(4)}} -  {{\phi^\prime }_{(5)}}
\\
\delta \phi_{{(5)}}  &=& -  \partial^{\mu }  {A^\prime_{{(2)} \mu }} - m  {{\phi^\prime }_{(8)}} + \frac{{1}}{2} {\phi^\prime }_{(2)}+ {{\phi^\prime }_{(4)}} + {{\phi^\prime }_{(5)}}
\\
\delta \phi_{{(6)}}  &=&   -2  {{\phi^\prime }_{(1)}} + {{\phi^\prime }_{(2)}}
\\
\delta \phi_{(7)}
&=&
\frac{d-3}{2} \phi^\prime_{(3)}
+T^\prime_\mu{}^\mu -\frac{1}{2}\phi^\prime_{(2)}
+\phi^\prime_{(4)} + \phi^\prime_{(11)}
\\
\delta \phi_{(8)} &=&{(\partial^2+m^2)}  {{\phi^\prime }_{(1)}} + {{\phi^\prime }_{(6)}}
\\
\delta \phi_{(9)} &=&   {(\partial^2+m^2)}  {{\phi^\prime }_{(3)}} - {{\phi^\prime }_{(6)}}
\\
\delta \phi_{(10)} &=&  {(\partial^2+m^2)}  {{\phi^\prime }_{(2)}} + 2 {{\phi^\prime }_{(6)}}
\end{eqnarray}
\vspace*{-6ex}
\begin{eqnarray}
%\\
\delta \phi_{(11)}  &=&   \partial^{\mu }  {A^\prime_{{(3)} \mu }} + m  {{\phi^\prime }_{(9)}} + {(\partial^2+m^2)}  {{\phi^\prime }_{(4)}} - {{\phi^\prime }_{(6)}}
\\
\delta \phi_{(12)}  &=&  -2 {{\phi^\prime }_{(7)}} + {{\phi^\prime }_{(8)}} + {{\phi^\prime }_{(10)}}
\\
\delta \phi_{(13)}
&=&
- \partial^\mu A^\prime_{(4)\mu}
- m \phi^\prime_{(4)}
- 2 m  \phi^\prime_{(11)}
+ \phi^\prime_{(9)}
\\
\delta \phi_{(14)} &=&  {(\partial^2+m^2)}  {{\phi^\prime }_{(7)}} + {{\phi^\prime }_{(9)}}
\\
\delta \phi_{(15)} &=&   -m  {{\phi^\prime }_{(5)}} + {(\partial^2+m^2)}  {{\phi^\prime }_{(8)}} + {{\phi^\prime }_{(9)}}
\\
\delta \phi_{(16)}  &=&     m  {{\phi^\prime }_{(5)}} + {(\partial^2+m^2)}  {{\phi^\prime }_{(10)}} + {{\phi^\prime }_{(9)}}
\\
\delta \phi_{(17)}
&=&
m \phi^\prime_{(7)}
- \frac{1}{2}\phi^\prime_{(1)}
+\phi^\prime_{(11)}
\\
\delta \phi_{(18)}
&=&
-m\phi^\prime_{(9)}
+ (\partial^2+m^2)\phi^\prime_{(11)}
+ \frac{1}{2}\phi^\prime_{(6)}
\\
\delta \phi_{(19)}&=&  m  {{\phi^\prime }_{(8)}} + m  {{\phi^\prime }_{(10)}} - \frac{{1}}{2} {\phi^\prime }_{(2)}+ 2 {{\phi^\prime }_{(11)}}
\end{eqnarray}

Let us turn to the gauge transformation for the gauge
parameters.

\subsubsection{The gauge transformations for the gauge parameters}
Let us decompose the vector (\ref{chi}) obeying the (\ref{state}) at
$n_1=n_2=1$ and having ghost number $-2$ as follows
\begin{eqnarray}
|\chi^{2}\rangle_{1,1} &=&
P_1^+P_2^+|\phi_1^{\prime \prime} \rangle_{0,0}
+P_1^+P_{12}^{G+}|A^{\prime \prime} \rangle_{1,0}
+P_{11}^+P_{12}^{G+}|\phi_2^{\prime \prime} \rangle_{0,0}\,,
\label{x-2}
\end{eqnarray}
where
\begin{eqnarray}
|A^{\prime \prime}\rangle_{1,0}  =
-i a_1^{+\mu} |0\rangle A^{\prime \prime}_\mu (x)
+ b_1^+ |0\rangle \phi^{\prime \prime}_{(3)} (x)
\,,
&\qquad&
|\phi^{\prime \prime}_{i}\rangle_{0,0} =
|0\rangle \phi^{\prime \prime}_{(i)}(x)
\,.
\end{eqnarray}

Substituting (\ref{x-1}), (\ref{x-2}) in the right relations
(\ref{dx0}) one gets the gauge transformations for the gauge
parameters
\begin{align}
&
\delta T_{\mu\nu}^\prime
=
\partial_{(\mu}  A^{\prime \prime}_{\nu) }
-\frac{1}{2}\eta_{\mu\nu}\phi^{\prime\prime}_{(2)}
,
&&
\delta A_{(1)\mu}^\prime  = - A^{\prime\prime}_\mu,
\\
&
\delta A_{(2)\mu}^\prime
=
\partial_\mu\phi^{\prime\prime}_{(1)} - A^{\prime\prime}_\mu,
&&
\delta A_{(3)\mu}^\prime
=
(\partial^2+m^2) A^{\prime\prime}_\mu,
\\
&
\delta A_{(4)\mu}^\prime
=
m A^{\prime\prime}_{\mu}
+\partial_{\mu}\phi^{\prime\prime}_{(3)},
&&
\delta H_\mu^\prime
=
-\partial_\mu \phi^{\prime\prime}_{(1)}
-A^{\prime\prime}_{\mu},
\\
&
\delta\phi_{(1)}^\prime = -\phi^{\prime\prime}_{(2)},
&&
\delta\phi_{(2)}^\prime  = - 2\phi^{\prime\prime}_{(2)},
\\
&
\delta\phi_{(3)}^\prime  =\phi^{\prime\prime}_{(2)},
&&
\delta\phi_{(4)}^\prime
=
-\partial^\mu A^{\prime\prime}_\mu - m\phi^{\prime\prime}_{(3)}
+\phi^{\prime\prime}_{(2)},
\\
&
\delta\phi_{(5)}^\prime
=
(\partial^2+m^2)\phi^{\prime\prime}_{(1)},
&&
\delta\phi_{(6)}^\prime
=
(\partial^2+m^2)\phi^{\prime\prime}_{(2)},
\\
&
\delta\phi_{(7)}^\prime
=
-\phi^{\prime\prime}_{(3)},
&&
\delta\phi_{(8)}^\prime
=
m \phi^{\prime\prime}_{(1)} - \phi^{\prime\prime}_{(3)},
\\
&
\delta\phi_{(9)}^\prime
=
(\partial^2+m^2)\phi^{\prime\prime}_{(3)},
&&
\delta \phi_{(10)}^\prime
=
- m \phi^{\prime \prime}_{(1)} - \phi ^{\prime \prime}_{(3)},
\\
&
\delta \phi_{(11)}^\prime
=
m\phi^{\prime\prime}_{(3)}
- \frac{1}{2}\phi^{\prime\prime}_{(2)},
\end{align}

\subsubsection{Gauge fixing and partial use of equations of motion}

Let us fix gauge symmetry completely by the gauge fixing
conditions (\ref{G1}) and (\ref{G2}) obtained at general
consideration.
It is easy to see that following fields will be eliminated
\begin{equation}
\phi_{(1)},\phi_{(2)},\phi_{(9)},\phi_{(13)},\phi_{(15)},\phi_{(17)},\phi_{(18)},\phi_{(19)},A_{(5)},A_{(6)},A_{(7)},H_{(2)},T_{(1)} \longrightarrow 0.
\end{equation}
Then using equations of motion for all fields but antisymmetric part
of the basic field $B_{[\mu\nu]}$, one sees that only $B_{[\mu\nu]}$
remains and Lagrangian for it, up to total derivative, is
\begin{eqnarray}
{\cal L}_{B_{[\mu\nu]}}
&=&
 B_{[ \mu \nu ]} (\partial^2+m^2) B^{[ \mu \nu ]} + 2 (\partial_{\mu} B^{[ \mu \lambda ]} ) ( \partial^{\nu} B_{[ \nu \lambda ]} )
 \\
&=& - \frac{1}{3} G_{\mu \nu \lambda} G^{\mu \nu \lambda}
+ m^2 B_{[ \mu \nu ]}B^{[ \mu \nu ]}
+\mbox{total derivative}
\end{eqnarray}
where we have defined field strength $G_{\mu \nu \lambda} $ of $B_{[
\mu \nu ]}$
\begin{eqnarray}
G_{\mu \nu \lambda}
:=
\partial_{\lambda} B_{[ \mu \nu ]}
+ \partial_{\mu} B_{[ \nu\lambda ]}
+ \partial_{\nu} B_{[ \lambda \mu ]}
.
\end{eqnarray}
Thus we have obtained the
gauge-invariant Lagrangian (\ref{L(1,1)}) for massive rank-2 antisymmetric
tensor field containing the complete set of auxiliary fields and
gauge parameters.

\section{Summary and discussion}

We have developed the general gauge invariant approach to Lagrangian
construction describing dynamics of massive bosonic higher spin
fields with index symmetry corresponding to two-row Young tableau
(\ref{basic}) in flat space-time of arbitrary dimension. The
obtained field model is a reducible gauge theory. The final
Lagrangian and gauge transformations are given by (\ref{L}) and
(\ref{dx0})--(\ref{dx5}) respectively. The Lagrangian automatically
contains the appropriate set of the St$\ddot{\rm u}$ckelberg fields
providing the gauge invariance of massive theory.

Generalization of this construction to the fields corresponding to
$k$-row Young tableau is straightforward. In this case one should
introduce $K$ pairs of creation and annihilation operators
(\ref{[]}), where $i,j=1,2,\ldots,k$. Constraint algebra contains
the generators $l_0$, $l_i$, $l_i^+$, $l_{ij}$, $l_{ij}^+$, $g_{ij}$
analogous to the case $k=2$.The new representation can be obtained
using the procedure developed in \cite{0505092}. BRST-BFV operators
$\tilde{Q}$ and $Q$ are defined straightforwardly and there will be
$k$ spin number operators $\sigma_i$ which commute with $Q$. The
eigenvalues of these operators are related with the spin of the
field like in the two-row case. Then Lagrangian and gauge
transformations are constructed analogously to the case $k=2$ which
was studied in the given paper.

\section*{Acknowledgements}
H.T. is thankful to high energy theory groups of CQUeST, Hiroshima
U. and KEK for discussions and kind hospitality. The work of I.L.B
and V.A.K was partially supported by the INTAS grant, project
INTAS-05-7928, the RFBR grant, project No.\ 06-02-16346 and grant
for LRSS, project No.\ 4489.2006.2. Work of I.L.B was supported in
part by the DFG grant, project No.\ 436 RUS 113/669/0-3 and joint
RFBR-DFG grant, project No.\ 06-02-04012. Work of V.A.K was
partially supported by the joint DAAD-Mikhail Lomonosov Programm
(Referat 325, Kennziffer A/06/16774).

\end{document}